\documentclass[a4paper]{article}

\usepackage{INTERSPEECH2021}

\usepackage{multirow}
\usepackage{enumerate}

\title{Exploring Deep Learning for Joint Audio-Visual Lip Biometrics}
\name{Meng Liu$^1$, Longbiao Wang$^{1,*}$, Kong Aik Lee$^{2,*}$, Hanyi Zhang$^1$, Chang Zeng$^3$, Jianwu Dang$^{1}$}
\address{$^1$College of Intelligence and Computing, Tianjin University, China \\
$^2$Institute for Infocomm Research, A$^\star$STAR, Singapore\\
$^3$National Institute of Informatics, Tokyo, Japan}

\email{liumeng2017@tju.edu.cn, longbiao\_wang@tju.edu.cn, kongaik.lee@gmail.com}

\begin{document}

\maketitle
\begin{abstract}
Audio-visual (AV) lip biometrics is a promising authentication technique that leverages the benefits of both the audio and visual modalities in speech communication. Previous works have demonstrated the usefulness of AV lip biometrics. However, the lack of a sizeable AV database hinders the exploration of deep learning-based audio-visual lip biometrics. To address this problem, we compile a moderate-size database using existing public databases. Meanwhile, we establish the DeepLip\footnote{https://github.com/DanielMengLiu/DeepLip} AV lip biometrics system realized with a convolutional neural network (CNN) based video module, a time-delay neural network (TDNN) based audio module, and a multimodal fusion module. Our experiments show that DeepLip outperforms traditional speaker recognition models in context modeling and achieves over 50\% relative improvements compared with our best single modality baseline, with an equal error rate of 0.75\% and 1.11\% on the test datasets, respectively. 
\end{abstract}
\noindent\textbf{Index Terms}: lip biometrics, audio-visual, speaker recognition, deep learning, visual speech

\section{Introduction}

Speaker recognition has rapidly developed in the past decades. Automatic speaker verification (ASV) systems play a crucial role in many applications, such as security access, e-commerce, teleworking and in-car systems. However, there is also increasing concern that the ASV systems are vulnerable to spoofing attacks, acoustically noisy environments, far field, and other complex multifaceted scenarios. In this regard, audio-visual (AV) biometrics \cite{aleksic2006audio, sadjadi20202019, chung2018voxceleb2} can be a viable solution. It is one of the most promising, user-friendly and low-cost biometrics that is resilient to spoofing \cite{aleksic2006audio}. The incorporation of additional modalities can alleviate problems of a single modality and improve system performance. So far, AV multimodal techniques have achieved good performance in speech recognition (lipreading) \cite{assael2016lipnet}, speech enhancement \cite{michelsanti2020overview}, speech separation \cite{ephrat2018looking} and emotion recognition \cite{huan2020video}. 

Among various AV biometrics technologies \cite{tao2020audio, qian2021audio, lu2019lip, wu2019lvid}, AV lip biometrics can be a foreseeably beneficial  approach in future authentication systems. First, AV lip biometrics focuses on the mouth region-of-interest (ROI).  The mouth ROI is tightly correlated to speech production since the lips, tongue, teeth and oral cavity are integral components of the articulator \cite{bothe1993visual}. Therefore, AV lip biometrics can be expected to extract correlated and complementary speaker characteristics between visual and audio modalities. Second, evidence from lipreading research \cite{wand2017improving, almajai2016improved} indicates that variation in the speakers identity causes performance degradation, revealing that lip sequences reflect substantial speaker characteristics. 
   
Since AV lip biometrics has been rarely investigated so far, relevant references need to be investigated by analogy from AV biometrics, lipreading, and speaker recognition. For audio modality, residual neural network (ResNet) \cite{qin2019far} and time-delay neural network (TDNN) systems \cite{snyder2017deep} have achieved remarkable performance for extracting deep speaker embedding. 
Since speech is a time series, 1D convolution-based TDNN better captures long-term temporal dependencies of speech signals than 2D convolution-based ResNet \cite{desplanques2020ecapa}. Considering a balance between high performance and light weight, the extended TDNN (E-TDNN) \cite{snyder2019speaker}, is a satisfactory framework for extracting audio-only speaker embedding. 

For visual modality, conventional visual speech feature representation include appearance-based and shape-based features \cite{aleksic2006audio}, which utilize lip geometry, parametric, or statistical models \cite{gomez2002biometric}. Traditional lip biometrics systems usually employed delicate manual features with a shallow statistics back-end model, e.g., Gaussian mixture model (GMM) and hidden Markov model (HMM) \cite{luettin1996speaker}. These approaches attain acceptable performance on small datasets, but the accuracy is still far from adequate for practical applications. Deep learning has outperformed traditional machine learning methods in most tasks. However, there is no deep-learning based AV lip biometrics, mainly because of the dataset constraint. With extensive data, we can establish the deep-learning based video-only system similar to those for lipreading. The usual lipreading framework uses a convolutional neural network (CNN) for front-end visual feature extraction and a recurrent neural network (RNN) for back-end model training. The deep learning method usually employs raw features (raw lip images) instead of the above manual features. Fully 2D convolution, fully 3D convolution \cite{assael2016lipnet} and a mixture of 2D and 3D convolution \cite{stafylakis2018deep, chen2020lipreading} were compared in \cite{chung2016lip}, and it was found that the latter can extract more discriminative deep features than 2D or 3D structures alone. Long short-term memory (LSTM) \cite{petridis2017end, stafylakis2018zero} , gated recurrent unit (GRU) \cite{petridis2018end} and temporal convolutional neural network (TCN) \cite{ma2020towards, martinez2020lipreading} models were designed for modeling the temporal dynamics of the sequence. TCN combined the RNN and CNN architecture for sequence modeling, and the training speed and performance outperformed the state-of-the-art lipreading system. 

In this paper, our contribution mainly focuses on the following aspects. Firstly, deep learning-based AV lip biometrics is limited due to the lack of adequate AV speaker data. We establish an audio-visual speaker database using public datasets and a deep-learning based baseline called DeepLip, which aggregates two well-performed systems in lipreading (visual-only) and speaker recognition (audio-only). Secondly, we show the effectiveness of applying deep learning to extract deep lip features, doing away laborious feature engineering. Thirdly, the fusion of speech and visual speech preliminarily proves that correlated and complementary information between visual speech and audible speech can be well utilized in person authentication. For multimodal fusion, \cite{aleksic2006audio, michelsanti2020overview, zhang2020multimodal} have discussed common approaches to realize the fusion of different modalities. The speaker embedding in AV lip biometrics consists of audio-only and video-only speaker embedding, which is an enhancement compared to the original single modality.


\section{Audio-Visual Modalities}

Audio-visual speaker recognition has attracted more and more attention recently. Some advanced generative adversarial networks (GAN) and talking head models can achieve favorable performance in generating fake videos. However, such videos still cannot mimic lip movements accurately. They may make the lip movements fit the text (cheat lipreading) but cannot cheat a lip biometrics system. Because lip movements are complicated, even if different people say the same word, their lip movements would vary due to the divergence of their articulator and pronouncing habits.

\begin{figure}[htbp]
	\centering
	\includegraphics[width=\linewidth]{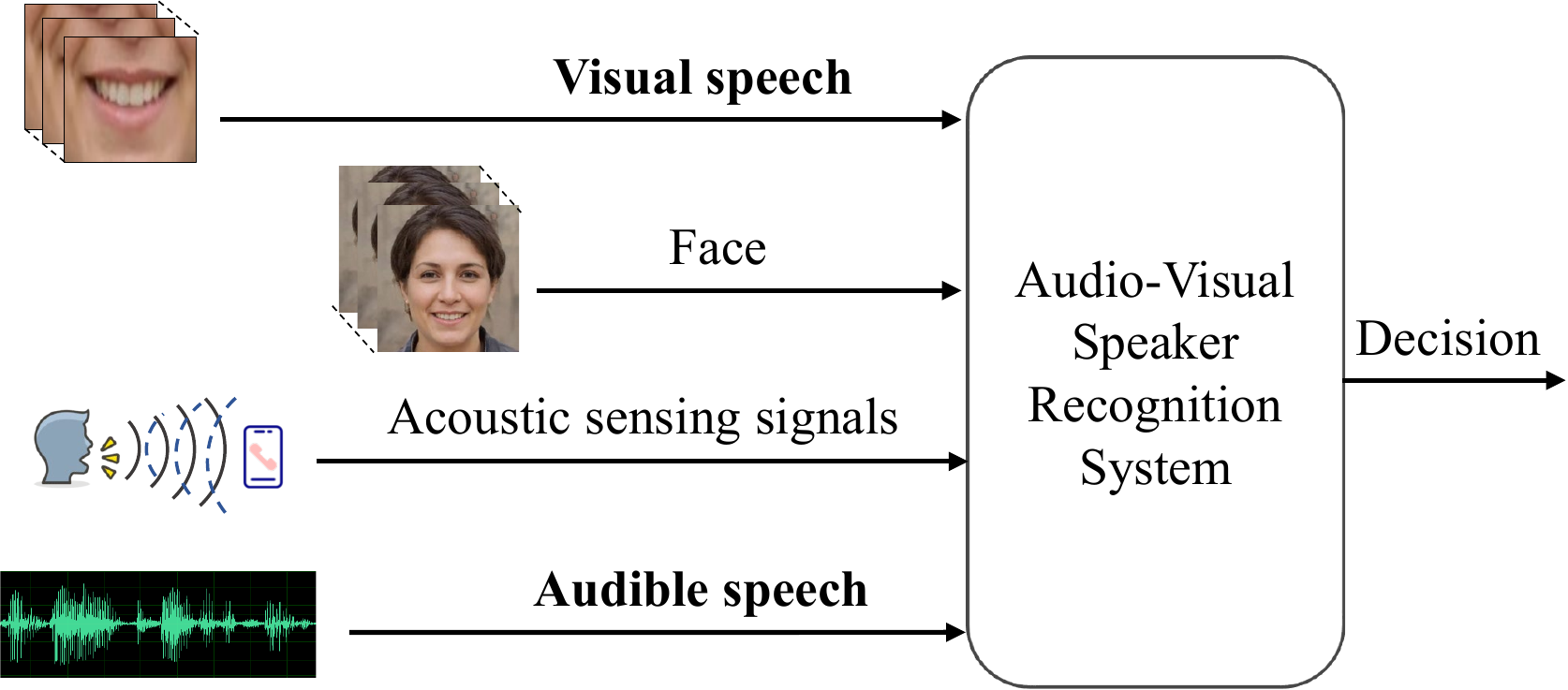}
	\caption{Various input modalities of audio-visual biometrics systems (Visual speech: lip movements).}
	\label{fig:AVbiometrics}
\end{figure} 

As shown in  Figure~\ref{fig:AVbiometrics}, existing works on AV speaker recognition mainly focuses on fusing facial features and audible speech \cite{tao2020audio, qian2021audio}, or facial features and acoustic signals collected by mobile acoustic sensing \cite{lu2019lip, wu2019lvid}. These methods have achieved promising improvements but have not explored hidden insights into the dynamics between audible speech and visual speech (lip movements). Moreover, \cite{wang2012physiological} defined physiological and behavioral lip biometrics by static lip images or dynamic lip videos. The study reported that dynamic lip shapes (movements) achieved better performance than static lip features (appearance), which also proved that visual speech (lip movements) contained substantial speaker characteristics. Therefore, lip biometrics cannot be replaced with face recognition, and it is more than just recognizing the appearance of lip. In this paper, our proposed DeepLip leverages both the visual speech and audible speech as the input of the ASV system.

\section{The DeepLip AV Lip Biometrics System}

\begin{figure*}[htbp]
	\centering
	\includegraphics[width=0.9\linewidth]{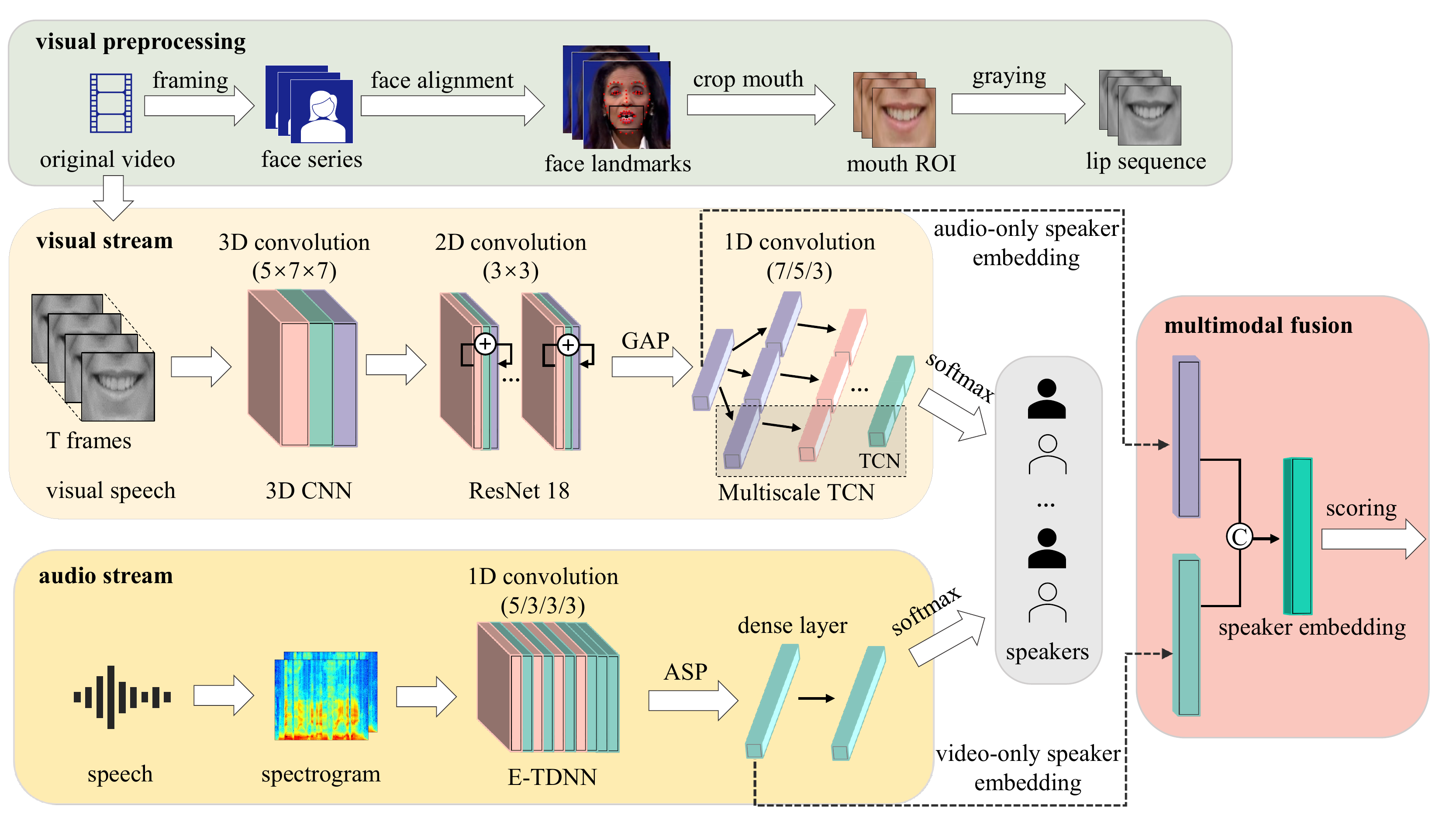}
	\caption{DeepLip: an overview of our deep-learning based lip biometrics architecture (GAP: global average pooling; ASP: attentive statistics pooling). }
	\label{fig:AVlipbiometrics}
\end{figure*}

Figure~\ref{fig:AVlipbiometrics} illustrates the overview of our deep-learning based lip biometrics architecture, referred to as DeepLip. The architecture consists of two streams: an audio-only stream to process audible speech, a visual-only stream to process the visual speech, and an audio-visual fusion module to fuse the audio and video speaker embeddings.

\subsection{Visual embedding using multi-stage CNN}

Preprocessing should be done to extract lip sequences. Each video sequence from the dataset is processed by 1) performing face detection and face alignment (obtaining landmarks), 2) aligning each frame to a reference mean face shape, 3) cropping a fixed 96 × 96 pixels ROI from the aligned face image, 4) transforming the cropped image  from RGB to gray level. 

Our visual stream uses a multi-stage convolutional neural network (MCNN) aggregating 3D, 2D and 1D convolutions, commonly used in state-of-the-art lipreading system \cite{martinez2020lipreading}. First, the grayscale lip sequences (a $B \times T \times H \times W$ tensor, corresponding to batch, frames, height and width) is segmented every 29 ms as CNN raw features, and a front-end 3D convolution with kernel size of 5 $\times$ 7 $\times$ 7 to extract visual feature. It is followed by an 18-layer residual network (2D convolution) feature encoder. A global average pooling is applied to obtain visual speech embedding ($B \times C \times T $, where $C$ is the output channel). Finally, a multiscale temporal convolutional network is used to model the time-indexed sequences. Several basic temporal convolutional blocks are stacked sequentially to act as a deep feature sequence encoder. The temporal receptive field of a standard TCN is kept fixed for all activations at a specific layer; multiple branches have been designed to provide variable receptive fields to fuse short-term and long-term temporal information during feature encoding. In the process, a variable-length augmentation procedure improves the generalization capabilities of the trained model when applied to sequences of varying lengths.

3D, 2D and 1D convolution play different roles during deep feature encoding. 3D convolution acts as a front-end feature extractor to extract fine-grained spatiotemporal information. 2D convolution in a deep residual network can further process image information in every frame and extract deep embedding. 1D non-casual temporal convolution is explicitly designed to capture short and long-term temporal dynamic lip movements.

\subsection{Audio embedding using Extended TDNN}

The TDNN (x-vector) based deep speaker embedding system has been widely recognized in large-scale speaker recognition tasks due to its stable and satisfactory performance. The extended TDNN x-vector architecture improves the performance over the original x-vector system introduced in \cite{snyder2019speaker}. Considering that our AV database is not large currently, it is enough to extract audio speaker embedding with the E-TDNN instead of state-of-the-art ECAPA-TDNN  \cite{desplanques2020ecapa}. In addition, E-TDNN is more appropriate due to its light-weight structure. The initial frame layers in E-TDNN consist of 1-dimensional dilated convolutional layers interleaved with dense layers. Every filter has access to all the features of the previous layer or input layer. The task of the dilated convolutional layers is to build up the temporal context gradually. After the statistics pooling, two fully connected layers are utilized, with the first one acting as a bottleneck layer to generate the low-dimensional audio speaker characterizing embedding. The output layer computes posterior probabilities for the training speakers.

\subsection{Audio-visual fusion}
Audio-visual fusion is of vital importance to prove the complementarity and relevance
between the audible speech and visual speech modalities. Early fusion, intermediate fusion and late fusion are three common approaches in multimodal integration \cite{aleksic2006audio, michelsanti2020overview}. Since the proposed DeepLip mainly acts as a baseline system to prove the complementarity of the audio and visual modalities, further fusion work will not be discussed in this work. Feature-level fusion and score-level fusion (late fusion) are implemented, respectively.

For feature fusion, audio speaker embedding and video speaker embedding form the speaker embedding via concatenation. L1 normalization is applied before concatenating. Fusion is implemented on the decision level using two independent streams. Each stream is regarded as an equal contributor to the final decision.


\section{DeepLip Database}

Dataset constraints are the major obstacles to the study of deep-learning based AV lip biometrics. We overcome this by combining suitable open-source datasets. Among  others we studied a number of audio-visual lipreading and biometrics datasets which include VoxCeleb2 \cite{chung2018voxceleb2}, LRW-1000 \cite{yang2019lrw}, LRW \cite{chung2016lip}, LRS, LRS2, LRS3 \cite{afouras2018lrs3}, GRID \cite{cooke2006audio}, LOMBARDGRID \cite{alghamdi2018corpus} and TCD-TIMIT \cite{harte2015tcd}. 

To construct the DeepLip database, our design principle is that an AV lip biometrics database must satisfy the following conditions:
\begin{itemize}
    \item Speaker labels. Most lipreading datasets ignore speaker labels, e.g., LRW, LRW-1000, LRS2 and LRS3.
    \item A sufficiently large number of speakers. Contrary to speech recognition, speaker recognition needs sufficient speakers instead of texts. 
    \item High visual resolution to extract mouth region. VoxCeleb2 failed because there are many stage lectures.
\end{itemize}

\begin{table}[htbp]
\caption{Three datasets used to construct the DeepLip database. Shown below are the released year, number of speakers, number of utterances, video resolution, and frames per second (FPS).}
\label{tab:datasets}
\centering
\setlength{\tabcolsep}{1mm}{
\begin{tabular}{rccrr}
\toprule
\textbf{Dataset} & \textbf{Year} &  \textbf{\# Spk.} & \textbf{\# Utt.}  & \textbf{Video Quality} \\
\midrule
GRID & 2006   & 34 & 34,000 & {360$\times$288,25fps} \\
LOMBARDGRID  & 2018   & 54 & 5,400 & {720$\times$480,25fps} \\
TCD-TIMIT & 2015  & 62 & 6,913 & {1920$\times$1080,30fps} \\
\bottomrule
\end{tabular}}
\end{table}

\begin{table}[htbp]
\caption{Partitions of the DeepLip database.}
\label{tab:partition}
\centering
\setlength{\tabcolsep}{1.5mm}{
\begin{tabular}{llcr}
\toprule
\textbf{Subset}    & \textbf{Source}  &  \textbf{\# Spk.}  &  \textbf{\# Utt.}\\
\midrule
training    & TCD-TIMIT &  62 & 6,913 \\
development    & LOMBARDGRID  &  18 & 1,774 \\
test1    & LOMBARDGRID  &  36  & 3,541 \\
test2    &  GRID    &  34 & 32,886 \\
\bottomrule
\end{tabular}}
\end{table}

Table~\ref{tab:datasets} shows the TCD-TIMIT, GRID and LOMBARDGRID used to construct the DeepLip database. The GRID and LOMBARDGRID corpora comprise phrases with fixed grammar, e.g., \emph{bin blue at A one now}, while the TCD-TIMIT corpus contains phonetically rich sentences, e.g., \emph{she had your dark suit in greasy wash water all year}. Table~\ref{tab:partition} shows the partitions of our DeepLip database. We removed the erroneous videos with lost frames or faces. Twenty thousand pairs of trials are randomly selected to form the test sets\footnote{https://github.com/
DanielMengLiu/DeepLip/database}.

\section{Experiments}

\subsection{Experimental setup}

The training of each stream is performed independently. For video stream training, we train for 300 epochs with a cosine scheduler, an initial learning rate of 0.05 and a weight decay of 1e-4, using cross entropy and Adam as the loss function and optimizer. We use the random crop of 88 $\times$ 88 pixels and random horizontal flip for data augmentation. We split the lip sequences into  several segments (every 29 frames corresponds to a segment, and segments less than 29 frames were abandoned). The basic TCN has four layers, and the multiscale TCN is composed of three branches with convolutional kernel sizes of 3, 5 and 7. 

For audio stream training, VoxCeleb1 and VoxCeleb2 datasets are pretrained for 30 epochs, and then we train for another 10 epochs on the training dataset.  We employ SGD and AM-Softmax \cite{liu2019large} as the optimizer and loss function, with an initial learning rate of 0.01, a weight decay of 1e-5 and a batch size of 256. The mel frequency cepstrum coefficient (MFCC) acts as the input speech feature, with the number of FFT points equal to 512 and the number of bins equal to 26. The configuration of E-TDNN is consistent with the classic setup described in \cite{snyder2019speaker}. Variable length augmentation is used in the training process. The universal background model (UBM) with a mixture of 64 is trained on the training set. Probabilistic linear discriminant analysis (PLDA) is trained on the development set with 150 principal components. Video and audio speaker embeddings are two 512 dimensional vectors.

\subsection{Audio-visual lip biometrics}

\begin{table}[htbp] 
	\caption{Performance comparison among audio-only, video-only and audio-visual systems using cosine similarity measurement. }
	\label{tab:fusion}
	\centering
	\begin{tabular}{lclc}
		\toprule
		\textbf{System Name} & \textbf{Testset} & \textbf{Integration} & \textbf{EER(\%)}              \\
		\midrule
		audio-only & test1 & single system & \textbf{1.98} \\	
		audio-only & test2 & single system & \textbf{2.12} \\
		video-only & test1 & single system & 5.73 \\
		video-only & test2 & single system & 8.65 \\	
		\midrule
		audio-visual & test1 & feature fusion & 0.84 \\
		audio-visual & test2 & feature fusion & \textbf{1.11} \\
		audio-visual & test1 & score fusion & \textbf{0.75} \\
		audio-visual & test2 & score fusion & 1.16 \\
		\bottomrule
	\end{tabular}
\end{table}

We enhance the audio-only system via transfer learning. To this end, VoxCeleb audio data and the training sets are used to pretrain and fine tune, respectively. Table~\ref{tab:fusion} shows a performance comparison among audio-only, video-only and audio-visual systems using cosine similarity measurement. The enhanced audio-only system can achieve an equal error rate (EER) of 1.98\% on test1 set and 2.12\% on test2 set, which serves as our best single system. Feature fusion, which simply concatenates the audio-only and video-only embeddings, significantly improves the performance of speaker recognition, with EERs of 0.84\% and 1.11\%. Remarkable complementary speaker information existed between visual speech and speech, revealing the potential to find a good feature fusion method. The score fusion for the audio-only and video-only systems also achieve a relative improvement of approximately 50\%. Above experimental results show that audio-visual lip biometrics can be one of the potential approaches in future speaker/person recognition.

\subsection{Single modality details}

A detailed single-system performance comparison between audio-only and video-only systems is shown in Table~\ref{tab:single}. We train all of the single systems on training set only and evaluate these models on our test1 and test2 sets. Two approaches are employed to measure the distance between the two test embeddings: a non-parametric method cosine similarity and a parametric model PLDA trained on our development set.

\begin{table}[t] 
	\caption{Performance comparison between traditional and deep-learning based audio-only and video-only systems. }
	\label{tab:single}
	\centering
	\begin{tabular}{lclcc}
		\toprule
		\textbf{Model} & \textbf{Testset} & \textbf{Measure} & \textbf{EER(\%)} & \textbf{minDCF}              \\
		\midrule
		GMM-UBM & test1 & - & 17.20 & 0.7688 \\
		GMM-UBM & test2 & - & 13.50 & 0.7558  \\
		E-TDNN & test1 & Cos & 12.63 & 0.5541 \\
		E-TDNN & test1 & PLDA & 8.16 & 0.5524 \\
		E-TDNN & test2 & Cos & 9.62 & 0.6351 \\
		E-TDNN & test2 & PLDA & 14.01 & 0.8856 \\
		\midrule
        MCNN & test1 & Cos & 5.73 & 0.2089 \\	
		MCNN & test1 & PLDA & 2.70 & 0.1126 \\
		MCNN & test2 & Cos & 8.65 & 0.3904 \\	
		MCNN & test2 & PLDA & 7.13 & 0.3371 \\
		\bottomrule
	\end{tabular}
\end{table}

The results demonstrates that the video stream (lip biometrics) achieves better performance than the audio stream, which proves that the visual speech does not contain fewer speaker characteristics than the audible speech. Compared with deep-learning model, the traditional speaker model GMM-UBM shows satisfactory performance on limited training data. PLDA significantly improves the performance on test1 set in both the video and audio streams. This is mainly because the development and test1 data are all from LOMBARDGRID dataset. The improvement decreases for the test2 data of the video-only stream and is even worse for the test2 data of the audio-only stream. It seems that PLDA is overfit for the LOMBARDGRID dataset, so we use cosine similarity as a measurement method in our audio-visual system.

\section{Conclusions and Future Work}
We have presented the DeepLip database that combines public datasets to promote the exploration of deep-learning based audio-visual lip biometrics. We also presented a deep-learning based audio-visual lip biometrics framework to leverage deep audio-visual speaker embeddings and their complementarity information.
Experimental results show the effectiveness of deep learning methods in AV lip biometrics. We obtained performance improvement over 50\% compared with that of our best single-modality system through simple score and feature fusion. The results prove the complementarity between audible speech and visual speech for AV lip biometrics.

With the convenience of audio-visual data collection in the future, we predict that audio-visual lip biometrics may soon become a popular research direction. More advanced techniques are expected to be introduced. This work serves as a baseline and proves the remarkable complementarity between deep audio and video speaker embeddings. It does not, however, focus on the correlation and alignment (multimodal integration) between them; this correlation will be explored in future research.


\section{Acknowledgements}
This work was supported by the National Key R\&D Program of China under Grant 2018YFB1305200, the National Natural Science Foundation of China under Grant 61771333 and the Tianjin Municipal Science and Technology Project under Grant 18ZXZNGX00330.

\bibliographystyle{IEEEtran}

\bibliography{mybib}

\begin{thebibliography}{10}
\providecommand{\url}[1]{#1}
\csname url@samestyle\endcsname
\providecommand{\newblock}{\relax}
\providecommand{\bibinfo}[2]{#2}
\providecommand{\BIBentrySTDinterwordspacing}{\spaceskip=0pt\relax}
\providecommand{\BIBentryALTinterwordstretchfactor}{4}
\providecommand{\BIBentryALTinterwordspacing}{\spaceskip=\fontdimen2\font plus
\BIBentryALTinterwordstretchfactor\fontdimen3\font minus
  \fontdimen4\font\relax}
\providecommand{\BIBforeignlanguage}[2]{{%
\expandafter\ifx\csname l@#1\endcsname\relax
\typeout{** WARNING: IEEEtran.bst: No hyphenation pattern has been}%
\typeout{** loaded for the language `#1'. Using the pattern for}%
\typeout{** the default language instead.}%
\else
\language=\csname l@#1\endcsname
\fi
#2}}
\providecommand{\BIBdecl}{\relax}
\BIBdecl

\bibitem{aleksic2006audio}
P.~S. Aleksic and A.~K. Katsaggelos, ``Audio-visual biometrics,''
  \emph{Proceedings of the IEEE}, vol.~94, no.~11, pp. 2025--2044, 2006.

\bibitem{sadjadi20202019}
S.~O. Sadjadi, C.~S. Greenberg, E.~Singer, D.~A. Reynolds, L.~Mason, and
  J.~Hernandez-Cordero, ``The 2019 nist audio-visual speaker recognition
  evaluation,'' \emph{Proc. Speaker Odyssey (submitted), Tokyo, Japan}, 2020.

\bibitem{chung2018voxceleb2}
J.~S. Chung, A.~Nagrani, and A.~Zisserman, ``Voxceleb2: Deep speaker
  recognition,'' \emph{Proc. Interspeech 2018}, pp. 1086--1090, 2018.

\bibitem{assael2016lipnet}
Y.~M. Assael, B.~Shillingford, S.~Whiteson, and N.~De~Freitas, ``Lipnet:
  End-to-end sentence-level lipreading,'' \emph{arXiv preprint
  arXiv:1611.01599}, 2016.

\bibitem{michelsanti2020overview}
D.~Michelsanti, Z.-H. Tan, S.-X. Zhang, Y.~Xu, M.~Yu, D.~Yu, and J.~Jensen,
  ``An overview of deep-learning-based audio-visual speech enhancement and
  separation,'' \emph{arXiv preprint arXiv:2008.09586}, 2020.

\bibitem{ephrat2018looking}
A.~Ephrat, I.~Mosseri, O.~Lang, T.~Dekel, K.~Wilson, A.~Hassidim, W.~T.
  Freeman, and M.~Rubinstein, ``Looking to listen at the cocktail party: a
  speaker-independent audio-visual model for speech separation,'' \emph{ACM
  Transactions on Graphics (TOG)}, vol.~37, no.~4, pp. 1--11, 2018.

\bibitem{huan2020video}
R.-H. Huan, J.~Shu, S.-L. Bao, R.-H. Liang, P.~Chen, and K.-K. Chi, ``Video
  multimodal emotion recognition based on bi-gru and attention fusion,''
  \emph{Multimedia Tools and Applications}, pp. 1--28, 2020.

\bibitem{tao2020audio}
R.~Tao, R.~K. Das, and H.~Li, ``Audio-visual speaker recognition with a
  cross-modal discriminative network,'' \emph{Proc. Interspeech 2020}, pp.
  2242--2246, 2020.

\bibitem{qian2021audio}
Y.~Qian, Z.~Chen, and S.~Wang, ``Audio-visual deep neural network for robust
  person verification,'' \emph{IEEE/ACM Transactions on Audio, Speech, and
  Language Processing}, 2021.

\bibitem{lu2019lip}
L.~Lu, J.~Yu, Y.~Chen, H.~Liu, Y.~Zhu, L.~Kong, and M.~Li, ``Lip reading-based
  user authentication through acoustic sensing on smartphones,'' \emph{IEEE/ACM
  Transactions on Networking}, vol.~27, no.~1, pp. 447--460, 2019.

\bibitem{wu2019lvid}
L.~Wu, J.~Yang, M.~Zhou, Y.~Chen, and Q.~Wang, ``Lvid: A multimodal biometrics
  authentication system on smartphones,'' \emph{IEEE Transactions on
  Information Forensics and Security}, vol.~15, pp. 1572--1585, 2019.

\bibitem{bothe1993visual}
H.-H. Bothe and F.~Rieger, ``Visual speech and coarticulation effects,'' in
  \emph{1993 IEEE International Conference on Acoustics, Speech, and Signal
  Processing}, vol.~5.\hskip 1em plus 0.5em minus 0.4em\relax IEEE, 1993, pp.
  634--637.

\bibitem{wand2017improving}
M.~Wand and J.~Schmidhuber, ``Improving speaker-independent lipreading with
  domain-adversarial training,'' \emph{Proc. Interspeech 2017}, pp. 3662--3666,
  2017.

\bibitem{almajai2016improved}
I.~Almajai, S.~Cox, R.~Harvey, and Y.~Lan, ``Improved speaker independent lip
  reading using speaker adaptive training and deep neural networks,'' in
  \emph{2016 IEEE International Conference on Acoustics, Speech and Signal
  Processing (ICASSP)}.\hskip 1em plus 0.5em minus 0.4em\relax IEEE, 2016, pp.
  2722--2726.

\bibitem{qin2019far}
X.~Qin, D.~Cai, and M.~Li, ``Far-field end-to-end text-dependent speaker
  verification based on mixed training data with transfer learning and
  enrollment data augmentation.'' in \emph{Interspeech}, 2019, pp. 4045--4049.

\bibitem{snyder2017deep}
D.~Snyder, D.~Garcia-Romero, D.~Povey, and S.~Khudanpur, ``Deep neural network
  embeddings for text-independent speaker verification.'' in
  \emph{Interspeech}, 2017, pp. 999--1003.

\bibitem{desplanques2020ecapa}
B.~Desplanques, J.~Thienpondt, and K.~Demuynck, ``Ecapa-tdnn: Emphasized
  channel attention, propagation and aggregation in tdnn based speaker
  verification,'' \emph{Proc. Interspeech 2020}, pp. 3830--3834, 2020.

\bibitem{snyder2019speaker}
D.~Snyder, D.~Garcia-Romero, G.~Sell, A.~McCree, D.~Povey, and S.~Khudanpur,
  ``Speaker recognition for multi-speaker conversations using x-vectors,'' in
  \emph{ICASSP 2019-2019 IEEE International Conference on Acoustics, Speech and
  Signal Processing (ICASSP)}.\hskip 1em plus 0.5em minus 0.4em\relax IEEE,
  2019, pp. 5796--5800.

\bibitem{gomez2002biometric}
E.~G{\'o}mez, C.~M. Travieso, J.~C. Brice{\~n}o, and M.~A. Ferrer, ``Biometric
  identification system by lip shape,'' in \emph{Proceedings. 36th Annual 2002
  International Carnahan Conference on Security Technology}.\hskip 1em plus
  0.5em minus 0.4em\relax IEEE, 2002, pp. 39--42.

\bibitem{luettin1996speaker}
J.~Luettin, N.~A. Thacker, and S.~W. Beet, ``Speaker identification by
  lipreading,'' in \emph{Proceeding of Fourth International Conference on
  Spoken Language Processing. ICSLP'96}, vol.~1.\hskip 1em plus 0.5em minus
  0.4em\relax IEEE, 1996, pp. 62--65.

\bibitem{stafylakis2018deep}
T.~Stafylakis and G.~Tzimiropoulos, ``Deep word embeddings for visual speech
  recognition,'' in \emph{2018 IEEE International Conference on Acoustics,
  Speech and Signal Processing (ICASSP)}.\hskip 1em plus 0.5em minus
  0.4em\relax IEEE, 2018, pp. 4974--4978.

\bibitem{chen2020lipreading}
X.~Chen, J.~Du, and H.~Zhang, ``Lipreading with densenet and resbi-lstm,''
  \emph{Signal, Image and Video Processing}, vol.~14, no.~5, pp. 981--989,
  2020.

\bibitem{chung2016lip}
J.~S. Chung and A.~Zisserman, ``Lip reading in the wild,'' in \emph{Asian
  Conference on Computer Vision}.\hskip 1em plus 0.5em minus 0.4em\relax
  Springer, 2016, pp. 87--103.

\bibitem{petridis2017end}
S.~Petridis, Z.~Li, and M.~Pantic, ``End-to-end visual speech recognition with
  lstms,'' in \emph{2017 IEEE International Conference on Acoustics, Speech and
  Signal Processing (ICASSP)}.\hskip 1em plus 0.5em minus 0.4em\relax IEEE,
  2017, pp. 2592--2596.

\bibitem{stafylakis2018zero}
T.~Stafylakis and G.~Tzimiropoulos, ``Zero-shot keyword spotting for visual
  speech recognition in-the-wild,'' in \emph{Proceedings of the European
  Conference on Computer Vision (ECCV)}, 2018, pp. 513--529.

\bibitem{petridis2018end}
S.~Petridis, T.~Stafylakis, P.~Ma, F.~Cai, G.~Tzimiropoulos, and M.~Pantic,
  ``End-to-end audiovisual speech recognition,'' in \emph{2018 IEEE
  international conference on acoustics, speech and signal processing
  (ICASSP)}.\hskip 1em plus 0.5em minus 0.4em\relax IEEE, 2018, pp. 6548--6552.

\bibitem{ma2020towards}
P.~Ma, B.~Martinez, S.~Petridis, and M.~Pantic, ``Towards practical lipreading
  with distilled and efficient models,'' in \emph{2018 IEEE international
  conference on acoustics, speech and signal processing (ICASSP)}.\hskip 1em
  plus 0.5em minus 0.4em\relax IEEE, 2021 (Accepted).

\bibitem{martinez2020lipreading}
B.~Martinez, P.~Ma, S.~Petridis, and M.~Pantic, ``Lipreading using temporal
  convolutional networks,'' in \emph{ICASSP 2020-2020 IEEE International
  Conference on Acoustics, Speech and Signal Processing (ICASSP)}.\hskip 1em
  plus 0.5em minus 0.4em\relax IEEE, 2020, pp. 6319--6323.

\bibitem{zhang2020multimodal}
C.~Zhang, Z.~Yang, X.~He, and L.~Deng, ``Multimodal intelligence:
  Representation learning, information fusion, and applications,'' \emph{IEEE
  Journal of Selected Topics in Signal Processing}, vol.~14, no.~3, pp.
  478--493, 2020.

\bibitem{wang2012physiological}
S.-L. Wang and A.~W.-C. Liew, ``Physiological and behavioral lip biometrics: A
  comprehensive study of their discriminative power,'' \emph{Pattern
  Recognition}, vol.~45, no.~9, pp. 3328--3335, 2012.

\bibitem{yang2019lrw}
S.~Yang, Y.~Zhang, D.~Feng, M.~Yang, C.~Wang, J.~Xiao, K.~Long, S.~Shan, and
  X.~Chen, ``Lrw-1000: A naturally-distributed large-scale benchmark for lip
  reading in the wild,'' in \emph{2019 14th IEEE International Conference on
  Automatic Face \& Gesture Recognition (FG 2019)}.\hskip 1em plus 0.5em minus
  0.4em\relax IEEE, 2019, pp. 1--8.

\bibitem{afouras2018lrs3}
T.~Afouras, J.~S. Chung, and A.~Zisserman, ``Lrs3-ted: a large-scale dataset
  for visual speech recognition,'' \emph{arXiv preprint arXiv:1809.00496},
  2018.

\bibitem{cooke2006audio}
M.~Cooke, J.~Barker, S.~Cunningham, and X.~Shao, ``An audio-visual corpus for
  speech perception and automatic speech recognition,'' \emph{The Journal of
  the Acoustical Society of America}, vol. 120, no.~5, pp. 2421--2424, 2006.

\bibitem{alghamdi2018corpus}
N.~Alghamdi, S.~Maddock, R.~Marxer, J.~Barker, and G.~J. Brown, ``A corpus of
  audio-visual lombard speech with frontal and profile views,'' \emph{The
  Journal of the Acoustical Society of America}, vol. 143, no.~6, pp.
  EL523--EL529, 2018.

\bibitem{harte2015tcd}
N.~Harte and E.~Gillen, ``Tcd-timit: An audio-visual corpus of continuous
  speech,'' \emph{IEEE Transactions on Multimedia}, vol.~17, no.~5, pp.
  603--615, 2015.

\bibitem{liu2019large}
Y.~Liu, L.~He, and J.~Liu, ``Large margin softmax loss for speaker
  verification,'' \emph{Proc. Interspeech 2019}, pp. 2873--2877, 2019.

\end{thebibliography}


\end{document}